\documentclass[12pt]{iopart}

\usepackage{hyperref}
\hypersetup{
	colorlinks=true,
	linkcolor=blue,
	filecolor=blue,      
	urlcolor=blue,
	citecolor=blue,
}
\usepackage{graphicx}  
\usepackage{subfig}
\usepackage{cite}
\usepackage{multirow}
\usepackage{makecell}
\usepackage{array}
\begin{document}

\title[]{Hadron-Quark phase transition in the context of GW190814}

\author{Ishfaq A. Rather$^1$, A. A. Usmani$^1$, S. K. Patra$^{2,3}$}

\address{$^1$Department of Physics, Aligarh Muslim University, Aligarh 202002, India}
\address{$^2$Institute of Physics, Bhubaneswar 751005, India}
\address{$^3$ Homi Bhabha National Institute, Training School Complex, Anushakti Nagar, Mumbai 400094, India}
\ead{ishfaqrather81@gmail.com}
\vspace{10pt}

\begin{abstract}
	The properties of the neutron stars are calculated for the hadronic matter within the density-dependent relativistic mean-field model (DD-RMF). The phase transition to the quark matter is studied and the hybrid star matter properties are systematically calculated using the Vector-Enhanced Bag model (vBag). The maximum mass of neutron star with DD-LZ1 and DD-RMF parameter sets is found to be around 2.55$M_{\odot}$ for pure hadronic phase and around 2$M_{\odot}$ for hadron-quark mixed phase using both Gibbs and Maxwell construction. The variation in tidal deformability for the hybrid EoS at 1.4$M_{\odot}$ depends upon the construction of phase transition. While it remains unchanged with Maxwell construction, but decreases with the increasing neutron star mass for Gibbs construction. The pure hadron matter EoS satisfies the maximum mass constraint from recently observed GW190814 data, implying a stiff neutron star EoS. The phase transition to quark matter satisfies the maximum mass and radius constraints from the GW170817 event. Therefore, we cannot exclude the possibility of the secondary object in GW190814 as a neutron star with a phase transition to the quark matter that satisfies the 2$M_{\odot}$ maximum mass limit.
	
\end{abstract}

%

\section{Introduction}
\label{intro}
The recent successful discovery of gravitational wave detection by LIGO and Virgo Collaborations (LVC) of a binary neutron star (BNS) merger GW170817 event \cite{PhysRevLett.119.161101,PhysRevLett.121.161101} has allowed us to study the dense matter properties at extreme conditions. The estimation of tidal deformability for neutron stars (NS) provided a new constraint on the NS Equation of State (EoS). The total mass of the GW170817 BNS merger was found to be around 2.7$M_{\odot}$ with the heavier component of 1.16-1.60$M_{\odot}$ for low spin priors and the maximum mass approached 1.9$M_{\odot}$ for high spin priors \cite{PhysRevX.9.011001}. The variation of tidal deformability with the radius as $\Lambda \propto R^5$ provides a strong constraint on the nuclear EoS at high density. After GW170817, the second possible BNS event GW190425 occurred with a total mass of 3.3-3.7$M_{\odot}$. The mass of its components with high spin prior are around 1.12-2.52$M_{\odot}$ \cite{Abbott_2020}. Recently, a new gravitational wave event reported by LVC as GW190814 \cite{Abbott_2020a} was observed with a 22.2-24.3$M_{\odot}$ black hole and 2.50-2.67$M_{\odot}$ secondary component. The secondary component attracted a lot of attention as it has no measurable tidal deformability signatures and no electromagnetic counterpart. The mass of the secondary component of GW190814 lies in the lower region of the so called mass-gap (2.5$M_{\odot}<M<5M_{\odot}$) which raises the question whether it is a light black hole or a supermassive NS. To explain the secondary component of GW190814, many interesting works have been proposed recently regarding its nature as super-massive NS, lightest black hole, or fast pulsar \cite{PhysRevC.103.025808,godzieba2020maximum,10.1093/mnrasl/slaa168,tan2020neutron,Zhang_2020,tsokaros2020gw190814,fattoyev2020gw190814,lim2020revisiting,tews2020nature}.\par

The mass, radius, and tidal deformability of a NS are determined by the EoS which is the relation between energy density and pressure. The value of NS maximum mass is assumed to be the most important parameter that determines the outcome of BNS mergers \cite{PhysRevD.73.064027,PhysRevLett.107.051102,PhysRevD.88.044026,PhysRevLett.111.131101,PhysRevD.92.044045,PhysRevD.94.024023,Lehner_2016,Radice_2018,K_ppel_2019}. Also, the proper knowledge of a NS maximum mass strongly constraints the poorly known EoS at several times the normal nuclear density \cite{PhysRevLett.105.161102,Hebeler_2013,doi:10.1146/annurev-nucl-102711-095018,Miller_2019,Annala2020}. Few measurements on the lower bounds of non-rotating NS properties like maximum are known accurately. The precise measurement of pulsar masses PSR J1614-2230 (1.928$\pm$0.017)$M_{\odot}$ \cite{Demorest2010},PSR J0348+0432(2.01$\pm$0.04)$M_{\odot}$ \cite{Antoniadis1233232}, and PSR J0740+6620 (2.04$^{+0.10}_{-0.09}$)$M_{\odot}$ \cite{Cromartie2020} show that the maximum mass of a NS should be at least 2$M_{\odot}$. The gravitational wave event GW170817 is interpreted as the possibility of an upper limit on the NS maximum mass which is around 2.3$M_{\odot}$ \cite{PhysRevD.100.023015,Rezzolla_2018,Margalit_2017}. With the new gravitational wave event GW190814 predicting a maximum mass in the range 2.5-2.67$M_{\odot}$, we see that the NS maximum mass is weakly constrained.\par 
Different models with different parameterizations have been used in the literature to construct the NS EoS at supranuclear densities. The Density functional theories (DFT) where the nucleon-nucleon interaction is effectively determined by fitting ground state properties of finite nuclei have been widely used to determine the saturation properties of high dense nuclear matter (NM) \cite{PhysRevC.5.626,SHEN1998435,PhysRevC.65.035802,refId0,PhysRevC.89.045807,PhysRevC.90.045802}.    
The NM EoS at saturation density from many-body theories are well constrained. The extrapolation of these EoSs to higher densities $\approx 4-5 \rho_0$, where $\rho_0$ is the nuclear saturation density, describe the properties of NSs. However, only few EoSs like NL3\cite{PhysRevC.55.540}and recently proposed BigApple \cite{fattoyev2020gw190814,das2020bigapple} generate massive NSs with maximum mass of $\approx$ 2.6$M_{\odot}$.\par 
The covariant density functional theory (CDFT) has been the most successful in explaining the finite as well as infinite NM properties. Walecka first proposed the Hartree approximation of CDFT, the $\sigma-\omega$ model \cite{Walecka:1974qa}, popularly known as the relativistic mean-field (RMF) model. It involves the basic mechanism where the nucleons interact through the meson exchange. The addition of $\rho$ meson, it's coupling to $\sigma$ and $\omega$, and the non-linear terms of $\sigma$-$\omega$ mesons improved the model and constrained the NM properties \cite{PhysRevLett.86.5647,SUGAHARA1994557,BOGUTA1977413,SEROT1979146}. The $\delta$ meson addition softened the EoS further and constrained the NS properties at high density regime \cite{PhysRevC.89.044001,Kumara:2017bti,PhysRevC.97.045806}. In the relativistic Hartree Fock model (RHF), the effect of the pion was taken into account \cite{PhysRevC.18.1510}. The self- and cross-coupling of various mesons can be replaced by the density-dependent nucleon-meson coupling constants in the density-dependent RMF (DD-RMF) models \cite{PhysRevLett.68.3408}. The density-dependent coupling constants allow a consistent calculation of neutron star and strange matter and yield results that are comparable with other models. It incorporates the properties of Dirac-Brueckner model using microscopic interactions at various densities as input. The extrapolation to higher densities is more constrained then for the phenomenological RMF calculations that use only information from the limited density range of finite nuclei for the determination of their parameters.

The most important difference from the RMF model is the contribution from the rearrangment term self energies to DD-RMF field equations. The rearrangment term accounts physically for static polarization effects in the nuclear medium. The contribution of rearrangment term  to  pressure implies that by not considering its contribution, it violates thermodynamic consistency because the mechanical pressure obtained from the energy-momentum tensor must coincide with the thermodynamic derivation. \par 
The DD-RMF and DD-RHF model parameters such as DD-ME1 \cite{PhysRevC.66.024306}, DD-ME2 \cite{PhysRevC.71.024312}, PKDD \cite{PhysRevC.69.034319}, generated massive NSs with a maximum in the range 2.3-2.5$M_{\odot}$, but the properties on a NS at the canonical mass, 1.4$M_{\odot}$ were not studied properly due to the limitation in the astronomical observations. Recently, several new DD-RMF parameter sets, DD-LZ1\cite{ddmex}, DD-MEX \cite{TANINAH2020135065}, and DDV, DDVT, DDVTD \cite{typel} were proposed. The latter parameter sets include the tensor couplings of the vector mesons to nucleons. All the parameter sets ensure that the causality is not violated. The large mass of NSs determined by these parameter sets can explain the secondary component of GW190814 as a NS. For symmetric nuclear matter, the softer EoSs like DDV, DDVT, and DDVTD completely satisfy the constraints from heavy-ion collisions \cite{Danielewicz1592} while as DD-LZ1 and DD-MEX produce stiff EoSs than the heavy-ion collision constraints. However, the constraints from heavy-ion collision are strongly model-dependent obtained from various inputs, such as $NN$ interactions which were analyzed adopting the RMF models. Hence, the EoSs considered cannot be completely outlined by constraints from heavy-ion collisions \cite{dexheimer2020future}.  \par 
The presence of exotic phases in the inner core of NSs has been studied over the past decade and the variation in the properties of NSs have constrained the NM EoS at high densities. The recent work \cite{Annala2020} has shown that the quarks are present in the NS core at several times the normal nuclear density. Hence the quark matter can exist inside the NSs in a deconfined phase \cite{PhysRevD.30.272,PhysRevD.30.2379} or as a mixed phase of hadrons and quarks (hybrid star) \cite{PhysRevD.46.1274,zel_2010,PhysRevD.88.085001,refId0,Bombaci2016}. Depending upon the phase transition  between outer hadronic matter and inner quark matter of the hybrid stars, the twin-star solution might appear as the mass-radius relation could exhibit two stable branches with same maximum mass but different radius \cite{Gomes:2018bpw, PhysRevD.99.103009,PhysRev.172.1325,2000A&A...353L...9G,Kampfer:1981yr}. A steep first-order phase transition (large density jump) combined with an incredibly stiff quark equation of state can generate twins. It's seen that when twin-star solution appears, the tidal deformability also displays two distinct branches with same maximum mass, which is different from the pure neutron and hybrid stars \cite{PhysRevD.98.083013,refId0,Burgio:2018yix}. A recent study shows that a NS with a maximum mass constraint of 2.5$M_{\odot}$ rules out twin star soultion \cite{PhysRevD.103.063042}.\par 
Several models have been proposed to study the hadron-quark phase transition in NSs like the simple MIT Bag model \cite{PhysRevD.30.2379,PhysRevD.9.3471,PhysRevD.17.1109} and the  Nambu-Jona-Lasinio (NJL) model \cite{KUBIS1997191,PhysRev.122.345,PhysRev.124.246,RevModPhys.64.649,BUBALLA2005205}, but there are limitations on their use regarding the hybrid stars (HSs) stability. The modification on the NJL model like 2+1 flavor NJL models has been successful in explaining the stable HSs \cite{PhysRevD.95.056018,PhysRevD.97.103013}. Furthermore, the modified models have bee successful in satisfying the recent constraints from GW170817.\par

The modified model of Bag model, Vector-Enhanced Bag model (vBag) \cite{Kl_hn_2015} was introduced as an effective model to study the astrophysical processes. It is favored over the simple bag model and NJL model as it accounts for the Dynamic Chiral Symmetry Breaking (D$\chi$SB) and also repulsive vector interactions. It also takes deconfinement into account for the construction of the mixed phase. The repulsive vector interaction is important as it allows the HSs to attain 2$M_{\odot}$ limit on the maximum mass \cite{universe4020030}.  The introduction of flavor-dependent chiral bag constants is motivated by fits to the pressure of the chirally restored phase. Furthermore, a deconfined bag constant is introduced in order to lower the energy per particle, thereby favoring stable strange matter. \par 
In this work, we use a few recently obtained DD-RMF parameter sets which generate a NS with a maximum mass around 2.5$M_{\odot}$, thus implying the nature of secondary component of GW190814 as a massive NS. Following a phase transition to the quark matter, we use its mass to put additional constraints on the NS maximum mass and on dense matter EoS.\par 

This paper is organized as follows: in section (\ref{sec:headings}), the density-dependent RMF model employed to study the NM is described followed by the vBag model for the quark matter. The phase transition from hadron matter to quark matter is also discussed. The equations governing the NS properties are explained in section (\ref{nsprop}). In section (\ref{results}), the parameter sets for the NM are defined. The star matter properties like EoS, mass, radii, and the tidal deformability for the given parameter sets are calculated. The properties of the phase transition are also studied. The summary and concluding remarks are finally given in section (\ref{summary}). \par

\section{Theory and Formalism}
\label{sec:headings}

\subsection{Nuclear Matter}
The Lagrangian density is the basic ansatz of the RMF theory where the nucleons interact through the exchange of mesons as Dirac particles. The mesons usually considered are scalar-isoscalar sigma $\sigma$, vector-isoscalar $\omega$, and vector-isovector $\rho$. However, the scalar-isoscalar $\delta$ meson is also considered to study the isovector effect on the scalar potential of the nucleon. NM properties like Symmetry energy and some high dense matter properties are affected by the contribution from $\delta$ meson \cite{KUBIS1997191,PhysRevC.89.044001}. The RMF Lagrangian contains the contribution from free baryon part and the meson part followed by the interaction between them along with the nucleon mass $M$, meson masses $m_{\sigma}$, $m_{\omega}$, $m_{\rho}$, $m_{\delta}$, and the corresponding coupling constants $g_{\sigma}$, $g_{\omega}$, $g_{\rho}$, and $g_{\delta}$ of the respective mesons. In DD-RMF model, the coupling constants vary with the denity i.e, density-dependent \cite{PhysRevLett.68.3408}. The coupling constants can be either dependent on the scalar density $\rho_s$ or the vector density $\rho_B$, but usually the vector density parameterizations is considered which influences only the self-energy instead of the total energy.\par
The DD-RMF Lagrangian is given as:


\begin{eqnarray}
\mathcal{L}  =\sum_{\alpha=n,p} \bar{\psi}_{\alpha} \Biggl\{\gamma^{\mu}\Bigg(i\partial_{\mu}-g_{\omega}(\rho_B)\omega_{\mu}
-\frac{1}{2}g_{\rho}(\rho_B)\gamma^{\mu}\rho_{\mu}\tau\Bigg) \nonumber \\
-\Bigg(M-g_{\sigma}(\rho_B)\sigma-g_{\delta}(\rho_B)\delta\tau\Bigg)\Biggr\} \psi_{\alpha} 
+\frac{1}{2}\Bigg(\partial^{\mu}\sigma \partial_{\mu}\sigma-m_{\sigma}^2 \sigma^2\Bigg) \nonumber \\
+\frac{1}{2}\Bigg(\partial^{\mu}\delta \partial_{\mu}\delta-m_{\delta}^2 \delta^2\Bigg)
-\frac{1}{4}W^{\mu \nu}W_{\mu \nu}
+\frac{1}{2}m_{\omega}^2 \omega_{\mu} \omega^{\mu}\nonumber \\
-\frac{1}{4}R^{\mu \nu} R_{\mu \nu}
+\frac{1}{2}m_{\rho}^2 \rho_{\mu} \rho^{\mu},
\end{eqnarray}

where $\psi$ denotes the nucleonic wave-function. $\sigma$, $\omega_{\mu}$, $\rho_{\mu}$ and $\delta$ represent the sigma, omega, rho, and delta meson fields, respectively. $g_{\sigma},g_{\omega},g_{\rho}$,and $g_{\delta}$ are the meson coupling constants which are density-dependent, and $m_{\sigma},m_{\omega},m_{\rho}$ and $m_{\delta}$ are the masses for  $\sigma, \omega, \rho$ and $\delta$ mesons respectively. The anti-symmetric tensor fields $W^{\mu \nu}$ and $R^{\mu \nu}$ are given by 
\begin{center}
	\begin{equation}
	W^{\mu \nu}=\partial^{\mu}W^{\nu}-\partial^{\nu}W^{\mu}
	\end{equation}
\end{center}
\begin{equation}
R^{\mu \nu}=\partial^{\mu}R^{\nu}-\partial^{\nu}R^{\mu}
\end{equation}
The density-dependent coupling constants are represented as:
\begin{equation}
g_i(\rho_B) = g_i(\rho_0) f_i(x)
\end{equation}
where,
\begin{equation}\label{eq5}
f_i(x) = a_i \frac{1+b_i (x+d_i)^2}{1+c_i(x+d_i)^2},i=\sigma,\omega
\end{equation}
is a function of $ x=\rho_B/\rho_{0}$ with $\rho_0$ is the NM saturation density.\par 
For the function $f_i(x)$, one has five constraint conditions $f_i(1)=1$,$f^{''}_{\sigma}(1)=f^{''}_{\omega}(1)$, $f^{''}_i(0)=0$ which reduce the number of free parameters from eight to three in eq.\ref{eq5}. The first two constraints lead to
\begin{equation}
a_i=\frac{1+c_i(1+d_i)^2}{1+b_i(1+d_i)^2},
3c_id_i^2=1
\end{equation} 
For $\rho$ and $\delta$ mesons, the coupling constants are given by an exponential dependence as
\begin{equation}
g_i(\rho_B)=g_i(\rho_0)exp[-a_i(x-1)]
\end{equation}

Following the Euler-Lagrange equation, we obtain equation of motion for nucleons and mesons as
\begin{equation}
\sum_{\alpha=n,p}\Bigg[i\gamma^{\mu}\partial_{\mu}-\gamma^0 \Bigg(g_{\omega}(\rho_B)\omega 	+\frac{1}{2}g_{\rho}(\rho_B)\rho \tau_3\\ +\sum_R (\rho_B)\Bigg)
-M_{\alpha}^*\Bigg]\psi_{i}=0,\\
\end{equation} 
\begin{center}
	
\begin{eqnarray}
	m_{\sigma}^2 \sigma = g_{\sigma}(\rho_B)\rho_s,\\
	m_{\omega}^2 \omega = g_{\omega}(\rho_B)\rho_B,\\
	m_{\rho}^2 \rho = \frac{g_{\rho}(\rho_B)}{2}\rho_3,\\
	m_{\delta}^2 \delta = g_{\delta}(\rho_B)\rho_{s3}.
	\end{eqnarray} 
\end{center}
$\sum_R$ is the rearrangment term introduced in the equation of motion of mesons due to the density dependent coupling constants.
\begin{equation}
\sum_R(\rho_B) = -\frac{\partial g_{\sigma}}{\partial \rho_B}\sigma \rho_s +\frac{\partial g_{\omega}}{\partial \rho_B}\omega \rho_B+\frac{1}{2}\frac{\partial g_{\rho}}{\partial \rho_B}\rho \rho_3-\frac{\partial g_{\delta}}{\partial \rho_B}\delta \rho_{s3},
\end{equation}
where $\rho_s$, $\rho_B$, $\rho_{s3}$, and $\rho_3$ are the scalar, baryon, and isovector densities, respectively, given by
\begin{equation}
\rho_s = \sum_{\alpha=n,p}\bar{\psi}\psi =\rho_{sp} +\rho_{sn}=\sum_{\alpha}\frac{2}{(2\pi)^3}\int_{0}^{k_{\alpha}}d^3k \frac{M_{\alpha}^*}{E_{\alpha}^*}
\end{equation}
\begin{equation}
\rho_B = \sum_{\alpha=n,p}\psi^{\dagger}\psi =\rho_{p} +\rho_{n}=\sum_{\alpha}\frac{2}{(2\pi)^3}\int_{0}^{k_{\alpha}}d^3k 
\end{equation}
\begin{equation}
\rho_{s3} = \sum_{\alpha}\bar{\psi}\tau_3\psi =\rho_{sp} -\rho_{sn}
\end{equation}
\begin{equation}
\rho_3 = \sum_{\alpha}\psi^{\dagger}\tau_3\psi =\rho_p -\rho_n
\end{equation}
The effective masses of nucleons are given as:
\begin{equation}
M_p^* =M-g_{\sigma}(\rho_B)\sigma -g_{\delta}(\rho_B)\delta
\end{equation},
and
\begin{equation}
M_n^* =M-g_{\sigma}(\rho_B)\sigma +g_{\delta}(\rho_B)\delta
\end{equation}
Also,
\begin{equation}
E_{\alpha}^*=\sqrt{k_{\alpha}^2+M_{\alpha}^{*2}},
\end{equation}
is the effective mass of nucleons with $k_{\alpha}$ as the nucleon momentum.
The energy-momentum tensor defined by the expression
\begin{equation}
T_{\mu \nu} = \sum_{i} \partial_{\nu}\psi_i \frac{\partial \mathcal {L}}{\partial (\partial ^{\mu} \psi_i)} - g_{\mu \nu} \mathcal{L},
\end{equation}
determines the energy density and pressure for the NM as
\begin{eqnarray} \label{eq18}
\mathcal{E}_H = \frac{1}{2}m_{\sigma}^2 \sigma^2-\frac{1}{2}m_{\omega}^2 \omega^2-\frac{1}{2}m_{\rho}^2 \rho^2+\frac{1}{2}m_{\delta}^2 \delta^2
+g_{\omega}(\rho_B)\omega \rho_B \nonumber \\
+\frac{g_{\rho}(\rho_B)}{2}\rho \rho_3 +\mathcal{E}_{kin},
\end{eqnarray}
\begin{eqnarray}\label{eq19}
P_H = -\frac{1}{2}m_{\sigma}^2 \sigma^2+\frac{1}{2}m_{\omega}^2 \omega^2+\frac{1}{2}m_{\rho}^2 \rho^2-\frac{1}{2}m_{\delta}^2 \delta^2 \nonumber \\
-\rho_B \sum_R (\rho_B)+P_{kin},
\end{eqnarray}
where, \\
\\
$\mathcal{E}_{kin}$ and $P_{kin}$ are the contributions to the energy density and pressure from the kinetic part,
\begin{eqnarray} \label{eq20}
\mathcal{E}_{kin} =\frac{1}{\pi^2}\int_{0}^{k_{\alpha}}k^2 \sqrt{k^2 +M_{\alpha}^{*2}} dk,\nonumber \\
P_{kin} = \frac{1}{3\pi^2}\int_{0}^{k_{\alpha}}\frac{k^4 dk}{\sqrt{k^2 +M_{\alpha}^{*2}}},
\end{eqnarray}

To determine the composition and properties of the NS matter, the $\beta$-equilibrium and charge neutrality are the two important conditions to be satisfied . The $\beta$-equilibrium condition for any baryon $B$ follows from $\mu_B = b_B \mu_n - q_B \mu_e$, where $\mu_B$ is the chemical potential with charge $q_B$ and baryon number $b_B$. For the present case, the $\beta$-equilibrium condition is given by therelation between chemical potential of nucleons and leptons as
\begin{equation}\label{c1}
\mu_e =\mu_{\mu} =  \mu_n - \mu_p. 
\end{equation}
where, 
\begin{eqnarray}
\mu_{\alpha=n,p} =\sqrt{k_{\alpha}^2 +M_{\alpha}^{*2}}+\Big[g_{\omega}(\rho_B)\omega +\frac{g_{\rho}(\rho_B)}{2}\rho \tau_3 +\sum_R (\rho_B)\Big],\nonumber \\
\mu_{l=\mu,e} = \sqrt{k_l^2 +m_l^2}.
\end{eqnarray}
The charge neutrality condition is given by
\begin{equation}
q_{total} = \sum_{i=n,p} q_i k_i^3/(3\pi^2)+\sum_l q_l k_l^3/(3\pi^2)=0,
\end{equation}
which implies, $\rho_p$ = $\rho_e+\rho_{\mu}$.

\subsection{Quark Matter}
The commonly used effective models to explain the presence of quark matter in NS cores either mimic quark confinement while keeping the quark masses constant like the Bag model \cite{PhysRevD.9.3471,PhysRevD.17.1109,PhysRevD.30.2379} or do exhibit the Dynamic Chiral Symmetry Breaking (D$\chi$SB) without confinement like Nambu-Jona-Lasino (NJL) models \cite{PhysRev.122.345,PhysRev.124.246,RevModPhys.64.649,BUBALLA2005205}. Both these types of models do not include the repulsive vector interactions, which is important in the study of NS properties as it allows the HSs to achieve 2$M_{\odot}$ limit which results from the recent constraints of PSR J1614-2230 \cite{Demorest2010}, PSR 0348+0432 \cite{Antoniadis1233232}, and PSR J0740+6620 \cite{Cromartie2020}. \\

In our recent work \cite{Rather_2020,Rather2020_1}, we have studied the phase transition between hadrons and quarks using the simple Bag model for the quark phase. In the present work, we employ an extension of the bag model, Vector-Enhanced Bag model (vBag) \cite{Kl_hn_2015} which is an effective model accounting for D$\chi$SB and repulsive vector interactions. It also accounts for the phenomenological correction to the quark matter EoS that describes the deconfinement and depends on the hadron EoS used to construct the phase transition.\\

The expression for the energy density and the pressure in vBag model are given as \cite{vbageos} 
\begin{equation}
\mathcal{E}_Q = \sum_{f=u,d,s} \mathcal{E}_{vBag,f}-B_{dc},
\end{equation}
\begin{equation}
P_Q = \sum_{f=u,d,s} P_{vBag,f}+B_{dc},
\end{equation}
where, $B_{dc}$ is the deconfined bag constant introduced to lower the energy per partile thereby favouring stable strange matter. $\mathcal{E}_{vBag,f}$
and $P_{vBag,f}$ are the energy density and pressure of a single quark flavor defined as:
\begin{equation}
\mathcal{E}_{vBag,f}(\mu_f) = \mathcal{E}_{FG,f}(\mu_f^*)+\frac{1}{2}K_{\nu}n_{FG,f}^2 (\mu_f^*)+B_{\chi,f},
\end{equation}   
\begin{equation}
P_{vBag,f}(\mu_f) = P_{FG,f}(\mu_f^*)+\frac{1}{2}K_{\nu}n_{FG,f}^2 (\mu_f^*)-B_{\chi,f},
\end{equation}
Here, FG represents the ideal, zero temperature Fermi gas formula. $K_{\nu}$ parameter is a coupling constant resulting from the vector interactions and controls the stiffness of the quark matter EoS \cite{Wei_2019}. $B_{\chi,f}$ represents the bag constant for a single quark flavor. The chemical potential $\mu_f^*$ of the system is parameterized by the relation
\begin{equation}
\mu_f =\mu_f^* +K_{\nu}n_{FG,f}(\mu_F^*).
\end{equation}  
An effective bag constant is defined in the vBag model so that the phase transition to quark matter occurs at the same chemical potential
\begin{equation}
B_{eff}=\sum_{f=u,d,s}B_{\chi,f}-B_{dc}.
\end{equation}
This allows us to illustrate how $B_{eff}$ can be used in case of two flavor and three flavor quark matter in HSs.\\

For the quark matter, the charge neutrality and $\beta$-equillibrium conditions are given as
\begin{equation}
\frac{2}{3}\rho_u -\frac{1}{2}(\rho_d+\rho_s)-\rho_e-\rho_u =0,
\end{equation}
\begin{equation}
\mu_s=\mu_d=\mu_u+\mu_e;
\mu_{\mu}=\mu_e.
\end{equation}
\subsection{Phase Transition}
As one moves from the outer crust to the inner core of a NS, the pressure and correspondingly the density increases, and the phase transition to the constituent parts of the nucleus take place-quarks. This phase transition ensures the softening of EoS, thereby reducing the NS maximum mass and tidal deformability. But this boundary between the pure NM and quark matter isn't well-defined \cite{PhysRevD.46.1274}. The $\beta$-equilibrium and charge neutrality conditions are generally the ones which detremine the density range over which a phase transition takes place. A few techniques have been used to construct the hadron-quark phase transition in NSs as defined in the refs.\cite{PhysRevD.46.1274,PhysRevC.60.025801,PhysRevC.75.035808,PhysRevC.66.025802,PhysRevC.89.015806}. The Gibbs construction (GC) \cite{PhysRevD.46.1274} and the Maxwell construction (MC) \cite{PhysRevD.88.063001} are widely used to describing the phase transition properties. The coexisting phase between hadrons and quarks exists over a finite range of pressure and density. In Maxwell construction, the transition to quark matter is determined by the local charge neutrality condition while as in Gibbs construction, the local charge neutrality is replaced by a global one which implies that both the hadron phase and quark phase are allowed to be separately charged. Furthermore, in GC, the pressure of the mixed-phase increases smoothly with the increasing density contrary to Maxwell construction, where the coexisting hadron and quark phases appear at the same pressure and baryon chemical potential but different electron chemical potential throughout the phase transition.\par

The global charge neutrality condition is given as
\begin{equation}
\chi \rho_{Q}+(1-\chi) \rho_{H}+\rho_l=0,
\end{equation} 
where $\chi= V_Q/(V_H+V_Q)$ is the quark volume fraction occupied by the quarks in the mixed phase which varies from $\chi=0$ in the pure hadron phase to $\chi=1$ in the pure quark phase. $\rho_Q$, $\rho_H$, and $\rho_l$ represent the quark, hadron phase and lepton charge densities, respectively.\par 
The pressure and chemical potential for the hadron-quark phase in GC is expressed as:
\begin{equation}
P_{H}(\mu_B,\mu_e) = P_{Q}(\mu_B,\mu_e) = P_{MP},
\end{equation}
and
\begin{equation}
\mu_{B,H} = \mu_{B,Q}; \mu_{e,H} = \mu_{e,Q}.
\end{equation}
The energy and the baryon density of the mixed phases are then given by:
\begin{equation}\label{e1}
\varepsilon_{MP} = \chi \varepsilon_{Q} +(1-\chi)\varepsilon_{H} +\varepsilon_l,
\end{equation}
and
\begin{equation}\label{e2}
\rho_{MP} = \chi \rho_{Q} +(1-\chi)\rho_{H}.
\end{equation}
Once the mixed phase is constructed, the eqs.(\ref{e1}) and (\ref{e2}) are solved to determine the properties of the mixed phase and eventually the structure of the star. \par 
In Maxwell Construction, the local charge neutrality condition is defined as:
\begin{equation}
\rho_H(\mu_B,\mu_e) =0; \rho_Q(\mu_B,\mu_e)=0.
\end{equation}
The expressions for the pressure and chemical potential
are then given as:
\begin{equation}
P_H(\mu_B,\mu_e)=P_Q(\mu_B,\mu_e) = P_{MP}
\end{equation}
\begin{equation}
\mu_{B,H} =\mu_{B,Q}
\end{equation}
The surface tension at the hadron-quark interface is not known precisely and hence both local as well as global charge neutrality conitions have been used for the mixed phase in the literature. In the present work, we employ both Gibbs and Maxwell methods to construct the phase transition between hadrons and quarks to determine the variations in the NS properties like mass, radius, and tidal deformability.

\section{Neutron star structure and properties}
\label{nsprop}
To determine the structure and properties of a spherical and static star, the Tolman Oppenheimer Volkoff equations coupled differential equations are solved \cite{PhysRev.55.364,PhysRev.55.374}
\begin{equation}\label{tov1}
\frac{dP(r)}{dr}= -\frac{[\mathcal{E}(r) +P(r)][M(r)+4\pi r^3 P(r)]}{r^2(1-2M(r)/r) }
\end{equation}
and
\begin{equation}\label{tov2}
\frac{dM(r)}{dr}= 4\pi r^2 \mathcal{E}(r)
\end{equation}
where $M(r)$ is the gravitational mass for a given choice of central energy density $\mathcal{E}_c$ and specific EoS. We have used $G=c=1$ here.\par
The tidal deformability $\lambda$ is defined in the linear order as the ratio of the induced quadrupole mass $Q_{ij}$ to the external tidal field $\mathcal{E}_{ij}$ as \cite{PhysRevD.81.123016,PhysRevC.95.015801}
\begin{equation}\label{l1}
\lambda=-\frac{Q_{ij}}{\mathcal{E}_{ij}} = \frac{2}{3}k_2 R^5
\end{equation}
The dimensionless tidal deformability follows from the $\lambda$ as
\begin{equation}\label{l2}
\Lambda=\frac{\lambda}{M^5}=\frac{2k_2}{3C^5}
\end{equation}
where $k_2$ is the second love number and $C=M/R$ is the compactness parameter. For the realistic star, both the love number and the radius for a stellar mass predicted by the NS EoS are fiexed with $k_2$ value of 0.05-0.15 \cite{Hinderer_2008}.The expression for the love number is given as \cite{PhysRevD.81.123016}
\begin{eqnarray}\label{l3}
k_2=\frac{8}{5}(1-2C)^2 [2C(y-1)]\Bigl\{2C(4(y+1)C^4
+(6y-4)C^3\nonumber \\
+(26-22y)C^2
+3(5y-8)C-3y+6)
-3(1-2C)^2(2C(y-1)\nonumber \\
-y+2)log\Big(\frac{1}{1-2C}\Big)\Bigr\}^{-1}.
\end{eqnarray}
The value of $y=y(R)$ can be computed by solving the following differential equation \cite{PhysRevC.95.015801,Hinderer_2008}
\begin{equation}\label{l4}
r\frac{dy(r)}{dr}+y(r)^2+y(r)F(r)+r^2 Q(r)=0,
\end{equation}
where,
\begin{equation}\label{l5}
F(r)=\frac{r-4\pi r^3 [\mathcal{E}(r)-P(r)]}{r-2M(r)},
\end{equation}
\begin{eqnarray}\label{l6}
Q(r)=\frac{4\pi r\Big(5\mathcal{E}(r)+9P(r)+\frac{\mathcal{E}(r)+P(r)}{\partial P(r)/\partial\mathcal{E}(r)}-\frac{6}{4\pi r^2}\Big)}{r-2M(r)}\nonumber \\
-4\Bigg[\frac{M(r)+4\pi r^3 P(r)}{r^2 (1-2M(r)/r)}\Bigg]^2.
\end{eqnarray}
With the given initial boundary conditions $P(0)=P_c$, $M(0)=0$, and $y(0)=2$ with $P_c$ as the central pressure to the surface of the star $P(R)=0$,$M(R)=M$, and $r(R)=R$, we solve above equations for a given central density to determine the mass, radius, love number, and the tidal deformability of a NS.

\section{Results and Discussions}
\label{results}
To study the hadron-quark phase transition and to determine the NS properties, several DD-RMF parameterizations are used. Recently, several new DD-RMF parameters were propsed, DD-MEX \cite{TANINAH2020135065}, DD-LZ1 \cite{ddmex}, and DDV, DDVT, DDVTD \cite{typel}. All these parameter sets were obtained by different groups by fitting the ground state properties of finite nuclei. These parameter sets include the necessary tensor couplings of the vector mesons to nucleons apart from the basic couplings. Apart from the above, we also used DD-ME1 \cite{PhysRevC.66.024306} and DD-ME2 \cite{PhysRevC.71.024312} parameter sets.\\

The nucleon and meson masses and the coupling constants between nucleon and mesons for DDV, DD-LZ1, DD-ME1, DD-ME2, and DD-MEX  parameter sets are shown in Table (\ref{tab1}). The independent parameters $a,b,c,d$ for $\sigma$, $\omega$, and $\rho$ mesons are also shown. None of the mentioned parameter sets in the table (\ref{tab1}) includes the contribution from delta meson and hence its mass and coupling constants are not shown here.\\

\begin{table}[hbt!]
	\centering
	\caption{Nucleon and meson masses and different coupling constants for various DD-RMF parameter sets. }
	\begin{tabular}{ cccccc }
		\hline
		\hline
		&DD-LZ1&DD-ME1&DD-ME2&DD-MEX&DDV \\
		\hline
		$m_n$ & 938.9000&939.0000&939.0000&939.0000&939.5654\\
		$m_p$&938.9000&939.0000&939.0000&939.0000&938.2721\\
		$m_{\sigma}$&538.6192&549.5255&550.1238&547.3327&537.6001\\
		$m_{\omega}$&783.0000&783.0000&783.0000&783.0000&783.0000\\
		$m_{\rho}$&769.0000&763.0000&763.0000&763.0000&763.0000\\
		$g_{\sigma}(\rho_0)$&12.0014&10.4434&10.5396&10.7067&10.1369\\
		$g_{\omega}(\rho_0)$&14.2925&12.8939&13.0189&13.3388&12.7704\\
		$g_{\rho}(\rho_0)$&15.1509&7.6106&7.3672&7.2380&7.8483\\
		\hline
		$a_{\sigma}$&1.0627&1.3854&1.3881&1.3970&1.2099\\
		$b_{\sigma}$&1.7636&0.9781&1.0943&1.3350&0.2129\\
		$c_{\sigma}$&2.3089&1.5342&1.7057&2.0671&0.3080\\
		$d_{\sigma}$&0.3799&0.4661&0.4421&0.4016&1.0403\\
		$a_{\omega}$&1.0592&1.3879&1.3892&1.3926&1.2375\\
		$b_{\omega}$&0.4183&0.8525&0.9240&1.0191&0.0391\\
		$c_{\omega}$&0.5386&1.3566&1.4620&1.6060&0.0724\\
		$d_{\omega}$&0.7866&0.4957&0.4775&0.4556&2.1457\\
		$a_{\rho}$&0.7761&0.5008&0.5647&0.6202&0.3326\\
		\hline
		\hline
	\end{tabular}
	\label{tab1}
\end{table}

It is necessary to mention that the coefficients of meson coupling constants $g_i,i=\sigma,\omega,\rho$ in DD-LZ1 parameter set are the values at zero density, while for other parameter sets, the values obtained are at nuclear saturation density ($\rho_0$). The parameter sets DDVT and DDVTD which contain the tensor coupling between vector mesons and nucleons are not considered in the present study as they predict an NS with a maximum mass less than 2 $M_{\odot}$. The addition of quarks in such hadronic EoS will soften the EoS and hence reduce the maximum mass further.\\

The symmetry energy parameter $J$ for the listed parameter sets are compatible with the $J=31.6\pm 2.66$ MeV obtained from various astrophysical observations \cite{LI2013276}. The symmetry energy slope parameter $L$ also satisfies the recent constraints $L=59.57\pm10.06$MeV \cite{PhysRevC.101.034303,DANIELEWICZ20141}. The currently accepted value of incompressibility determined from the isoscalar giant monopole resonance (ISGMR) lies in the range $K_0=240\pm20$MeV. The $K_0$ value for all the given parameter sets satisfies this range except for the DD-MEX which predicts a little higher value.\par

\begin{table}[hbt!]
	\centering
	\caption{NM properties Binding energy ($E/A$), incompressibility ($K_0$), symmetry energy ($J$), slope parameter ($L$) at saturation density for various DD-RMF parameter sets. }
	\begin{tabular}{ cccccc }
		\hline
		\hline
		&DD-LZ1&DD-ME1&DD-ME2&DD-MEX&DDV \\
		\hline
		$\rho_0(fm^{-3})$ &0.158&0.152&0.152&0.152&0.151\\
		$E/A$(MeV)&-16.126&-16.668&-16.233&-16.140&-16.097\\
		$K_0$(MeV)&231.237&243.881&251.306&267.059&239.499\\
		$J$(MeV)&32.016&33.060&32.310&32.269&33.589\\
		$L$(MeV)&42.467&55.428&51.265&49.692&69.646\\
		$M_n^*/M$&0.558&0.578&0.572&0.556&0.586\\
		$M_p^*/M$&0.558&0.578&0.572&0.556&0.585\\
		\hline
		\hline
	\end{tabular}
	\label{tab2}
\end{table}
\vspace{0.5cm}
\begin{figure}[hbt!]
	\centering
	\includegraphics[width=10cm, height=8cm]{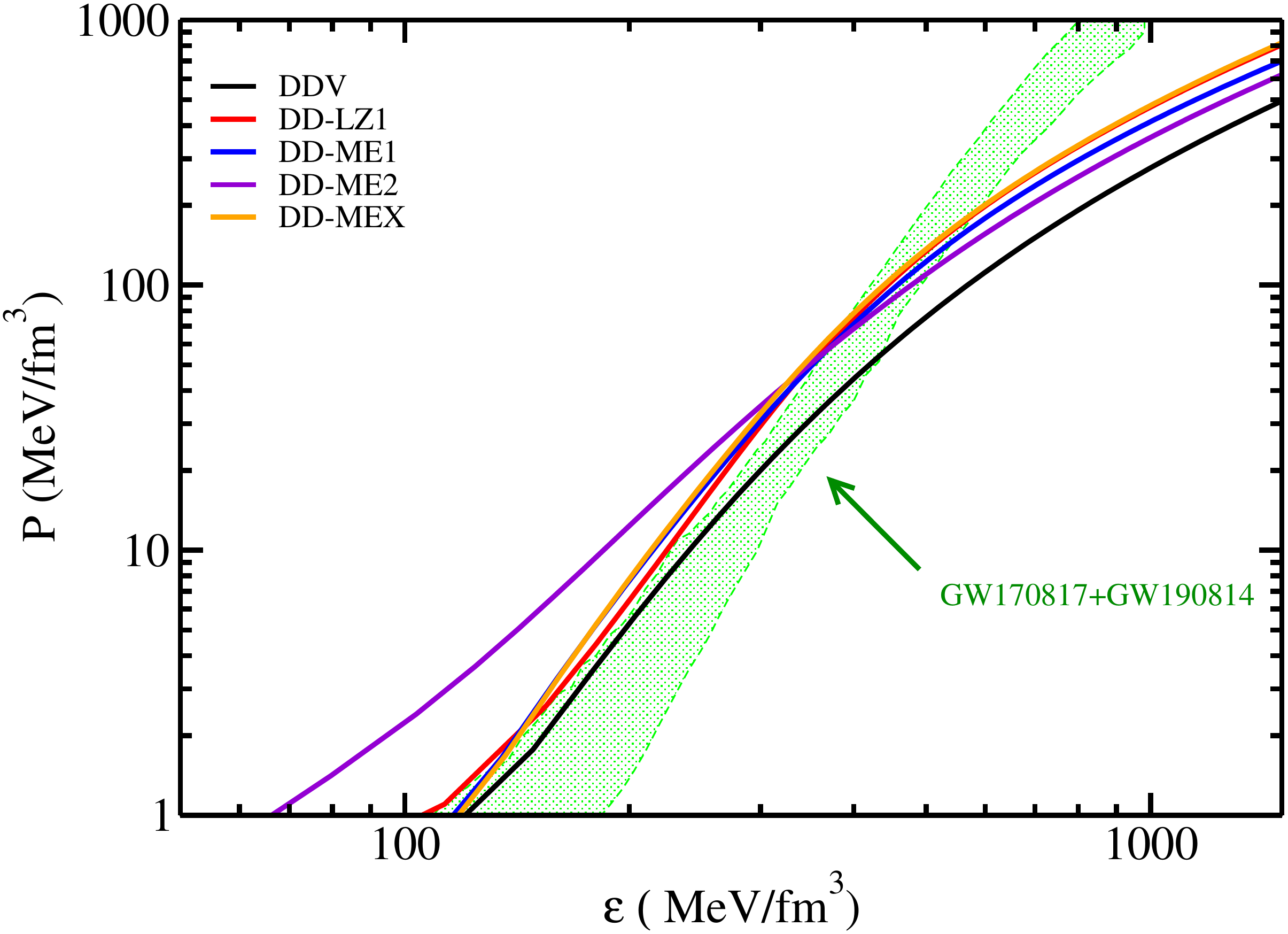}
	\caption{(color online) Energy density vs pressure for the given DDV, DD-LZ1, DD-ME1, DD-ME2, and DD-MEX parameter sets. The joint constraints from GW170817 and GW190814 shown are taken from \cite{Abbott_2020a}. }
	\label{fig1} 
\end{figure}
Fig.\ref{fig1} shows the variation of pressure with energy density (EoS) for an NS in beta-equilibrium and charge-neutral conditions. The DD-ME2 parameter set produces stiff EoS at low densities while DD-LZ1 and DD-MEX produce stiff EoS in high-density region. DDV set produces soft EoS at both low and high densities and hence defines an NS with low maximum mass as compared to other sets. The stiff EoS produced by the given parameter sets results in a high pressure due to strong vector potentials. The recent constraints on the EoS from GW170817 and GW190814 are also shown in the shaded area. The joint constraints were calculated by assuming a spectral distribution of EoS conditioned in GW170817 and re-weighted each EoS by the probability that its maximum mass is at least as large as the secondary component of GW190814. This was introduced by considering the GW190814 event as NS-Black hole (NSBH) merger, with its secondary component assumed to be a NS. For this scenario, the  maximum mass should be not less than secondary component of GW190814, which constraints the distribution of EoSs compatible with astrophysical data \cite{Abbott_2020a}.  With energy density less than $\mathcal{E} \approx$ 600 MeV/fm$^3$, the DD-RMF EoS's satisfy the constraints from the gravitational waves. As the energy density increases, the pressure from the obtained EoS's starts to lower than the gravitational wave constraints. For the unified EoS, the Baym-Pethick-Sutherland (BPS) EoS \cite{Baym:1971pw} is used for the outer crust part. For the inner crust, the EoS in the non-uniform matter is generated by using DD-ME2 prameter set in Thomas-Fermi approximation \cite{PhysRevC.79.035804,PhysRevC.94.015808,rather2020effect}.  

To construct the hadron-quark phase and determine the phase transition properties of a three-flavor configuration, we will consider the effective bag constant with a value of $B_{eff}^{1/4}$=130 MeV and $B_{eff}^{1/4}$=160 MeV. The value of $K_\nu$
parameter varies for two flavor and three flavor configurations. In the present study, we  will keep its value fixed at $K_{\nu} = 6$ GeV$^{-2}$.

\begin{figure}[hbt!]
	\centering
	\includegraphics[width=10cm,height=8cm]{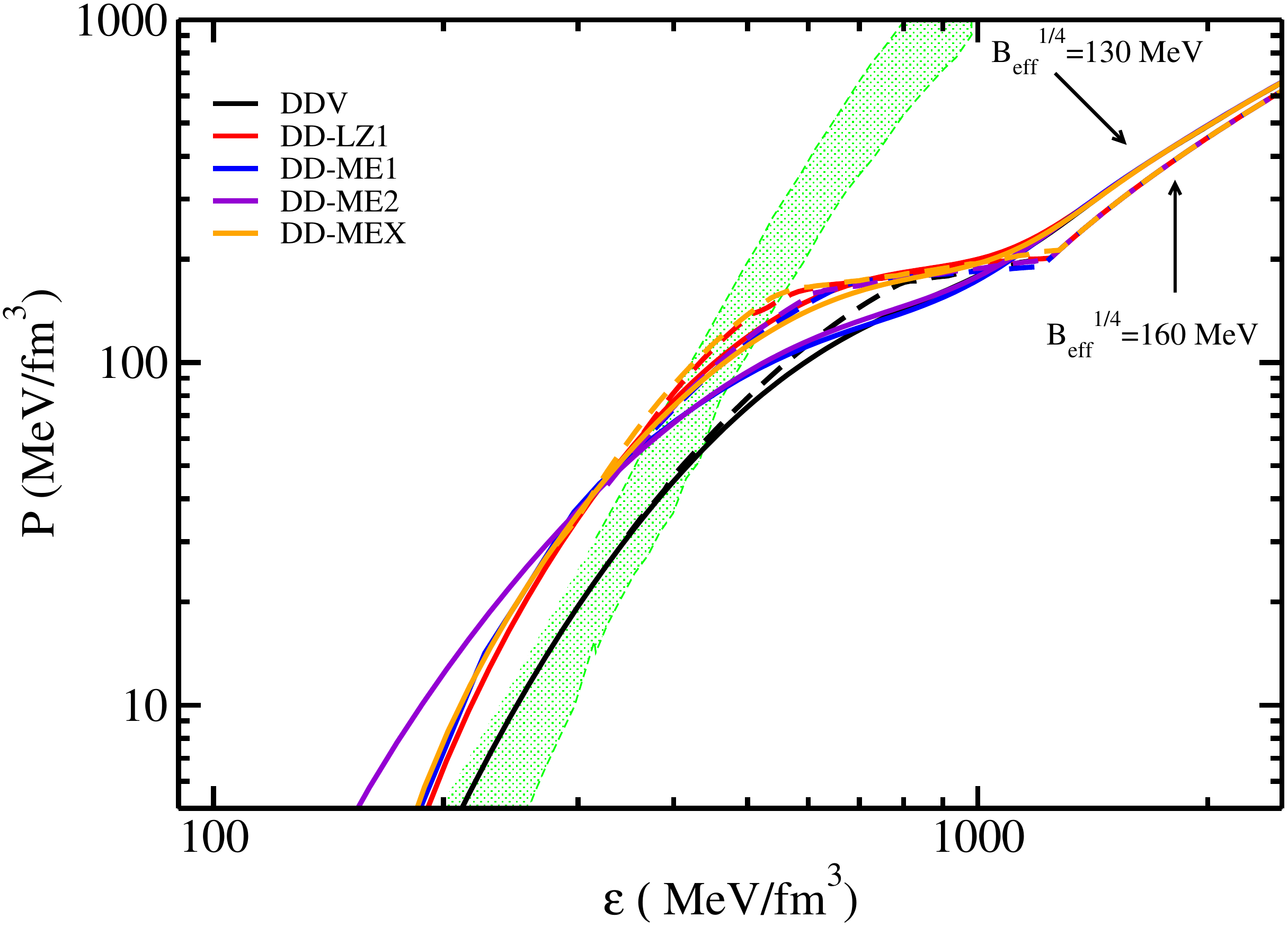}
	\caption{(color online) Equation of state for the hadron-quark phase transition for different DD-RMF hadronic parameter sets and three flavor quark matter at $B_{eff}^{1/4}$=130 \& 160 MeV using Gibbs construction. The solid (dashed) lines represent the phase transition at $B_{eff}^{1/4}$=130 MeV ($B_{eff}^{1/4}$=160 MeV). }
	\label{fig2} 
\end{figure}

Fig. (\ref{fig2}) shows the hadron-quark phase transition
with DD-RMF parameter sets for hadronic matter and vBag model for quark matter using the Gibbs method for constructing mixed-phase. The global charge neutrality condition ensures a smooth transition between the two phases. For effective bag constant $B_{eff}^{1/4}$=160 MeV, the phase transition takes place at higher density as compared to the bag value $B_{eff}^{1/4}$=130 MeV. In DDV EoS, the transition to quark matter at $B_{eff}^{1/4}$=130 MeV starts from the energy density $\mathcal{E} \approx$ 400MeV/fm$^3$ and ends at around $\mathcal{E} \approx$ 1200 MeV/fm$^3$ which corresponds to the density range $\rho_B =(2.47-4.03)\rho_0$. For $B_{eff}^{1/4}$=160 MeV, the phase transition region exists from $\rho_B =(3.69-5.31)\rho_0$. Similarly for DD-MEX EoS, the phase transition begins from 2.45 to 4.44$\rho_0$ and 3.09 to 5.57$\rho_0$ for bag values 130 and 160 MeV, respectively. It is clear that the phase transition for higher bag values occurs at higher densities and with large mixed-phase region.\par

\begin{figure}[hbt!]
	\centering
	\includegraphics[width=10cm, height=8cm]{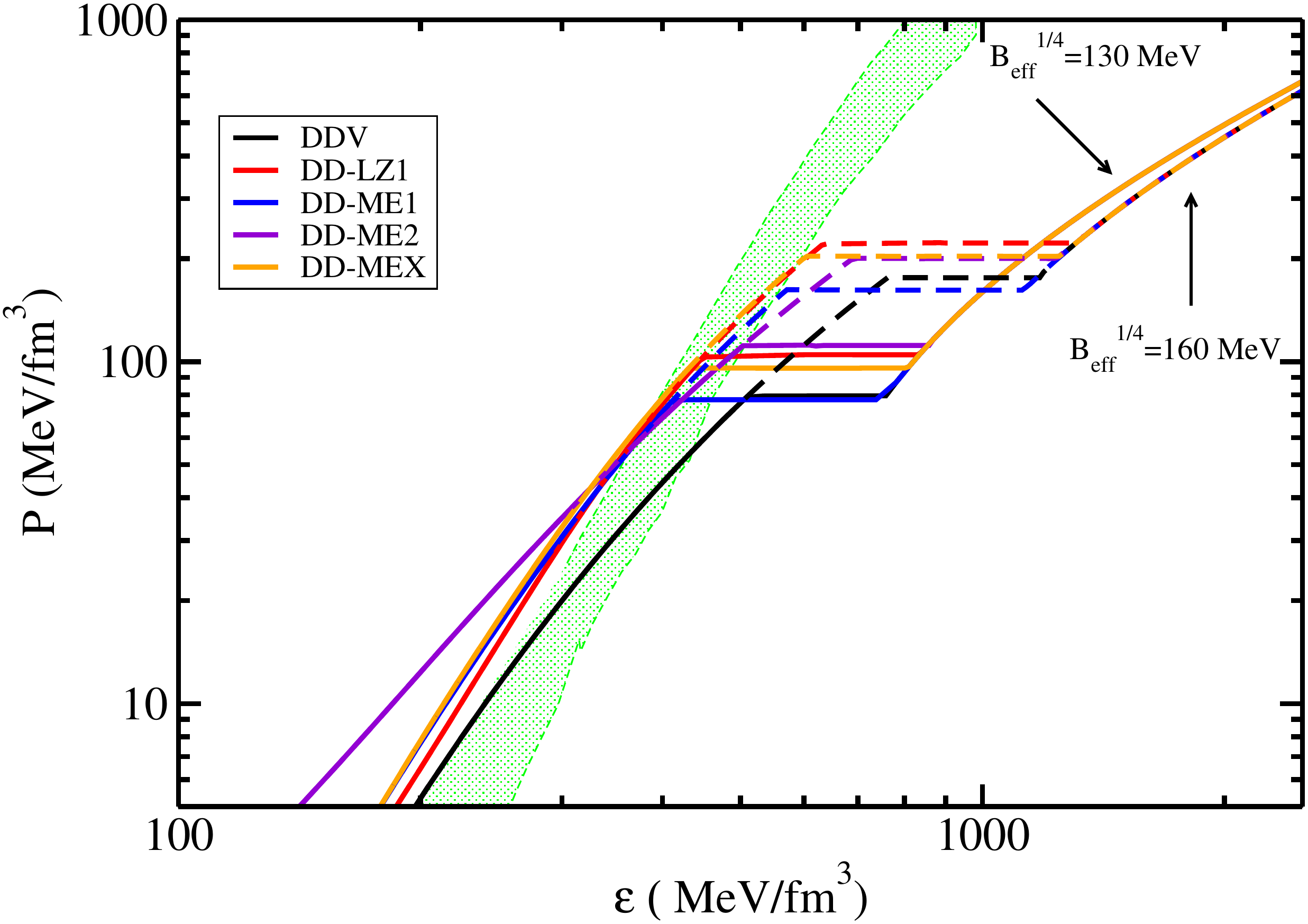}
	\caption{(color online) Same as fig. (\ref{fig2}) but using Maxwell construction. }
	\label{fig3} 
\end{figure}

Fig. (\ref{fig3}) represents the hadron-quark phase transition using Maxwell construction method. The local charge neutrality condition allows the phase transition to take place at constant pressure which results in a sharp shift from hadron matter to quark matter. For DDV EoS, the phase transition occurs in the density region $(2.62-3.91)\rho_0$ for $B_{eff}^{1/4}$=130 MeV and $(4.08-5.23)\rho_0$ for bag constant $B_{eff}^{1/4}$=160 MeV. As clear from Fig. \ref{fig2} and \ref{fig3}, the EoS at low densities (hadronic) satisfies the joint GW constraints. The phase transition to quark matter satisfies this constraint at beginning of the mixed phase for low value of bag constant. As the pure quark phase begins, the softness of EoS shifts away from the joint GW170817 and GW190814 constraints. This implies that a too stiff EoS is required for a quark matter to satisfy these constraints. 
\begin{table}
	\centering 
	\caption{Phase transition density for hadron-quark matter at $B_{eff}^{1/4}$=130 \& 160 MeV using both Maxwell and Gibbs construction methods. $\rho_{MP}$ represents density of the mixed phase region in terms of the saturation density $\rho_{0}$ which has the dimensions of fm$^{-3}$. }

	\begin{tabular}{c|cccc|cccc}
		\hline
		&&&&\multicolumn{2}{c}{$\rho_{MP}(\rho_0)$}& \\
		\hline
		\multirow{3}{*}{EoS} & \multicolumn{4}{c|}{Gibbs Construction} & %
		\multicolumn{4}{c}{Maxwell Construction}\\
		\cline{2-9}
		& \multicolumn{2}{c}{130 MeV} & \multicolumn{2}{c|}{160 MeV} & \multicolumn{2}{c}{130 MeV} & \multicolumn{2}{c}{160 MeV} \\

		\hline
		DDV&2.47-4.03&&3.69-5.31&&2.62-3.91&&4.08-5.23& \\
		
		DD-LZ1&2.56-4.23 &&3.04-5.43 &&2.71-4.18 &&3.21-5.24& \\
		
		DD-ME1&2.70-4.18&&3.04-5.47&&2.89-4.11&&3.39-5.41& \\
		
		DD-ME2& 2.87-4.39&&3.42-5.53&&3.01-4.35&&3.75-5.44&\\
		
		DD-MEX&2.45-4.44&&3.09-5.57&&2.49-4.28&&3.47-5.49& \\
		\hline
	\end{tabular}
	\label{tab3}
\end{table}

Table (\ref{tab3}) shows the phase transition density region between hadron and quark matter at bag values $B_{eff}^{1/4}$=130 and 160 MeV using both Maxwell and Gibbs construction. It is clear that the mixed phase region exists between $(2-6)\rho_0$, where $\rho_0$ is the nuclear saturation density, which is important in obtaining NSs with maximum mass larger than 2 $M_{\odot}$ \cite{PhysRevC.66.025802,Masuda_2013,10.1093/ptep/ptt045}. The increase in the value of bag constant delays the phase transition and softens the pure quark phase \cite{PhysRevD.88.063001,Bhattacharyya_2010,Lenzi_2012} as seen in fig's. (\ref{fig2},\ref{fig3}). Also, the phase transition in case of GC starts earlier than MC which is consistent with the work from ref's \cite{Bhattacharyya_2010,PhysRevD.85.023003}. However, the width of mixed phase region in GC is much broader than in MC and it increases further for GC as the bag constant increases. These properties certainly affect the  mass and radius.

\begin{figure}[hbt!]
	\centering
	\includegraphics[width=14cm,height=10cm]{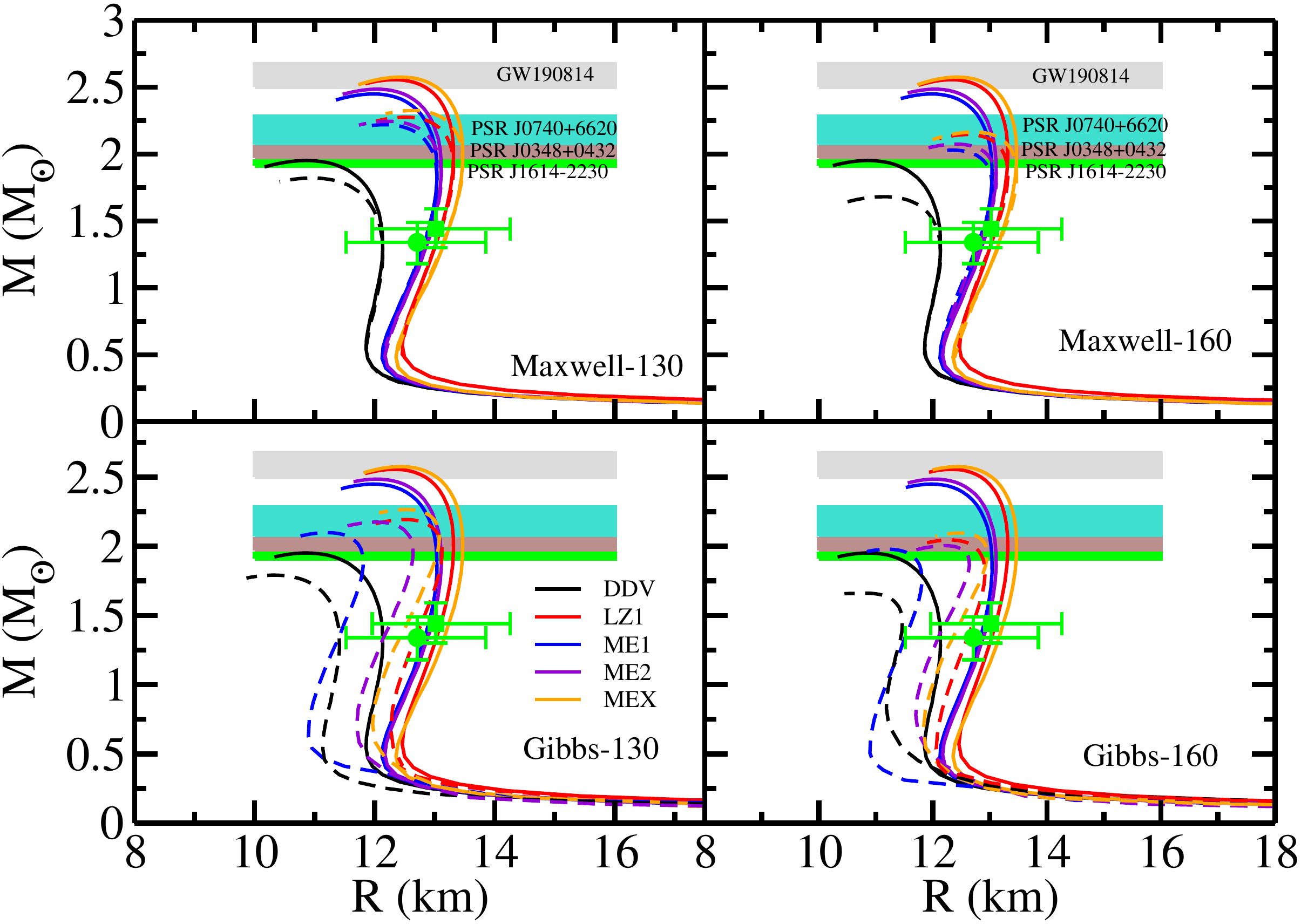}
	\caption{(color online) Mass-Radius profile for pure hadronic DD-RMF parameters and HSs for different bag constants. The solid (dashed) lines represent MR plot for pure hadronic matter (hybrid NSs). The upper panels display the HS configuration with MC and the lower panels represent the same configuration with GC. The recent constraints on the mass \cite{Abbott_2020a,Demorest2010,Antoniadis1233232,Cromartie2020} and on the radii from NICER's observation \cite{Miller_2019a,Riley_2019} are also shown.}
	\label{fig4}
\end{figure}

\begin{figure}[hbt!]
	\centering
	\includegraphics[width=14cm,height=10cm]{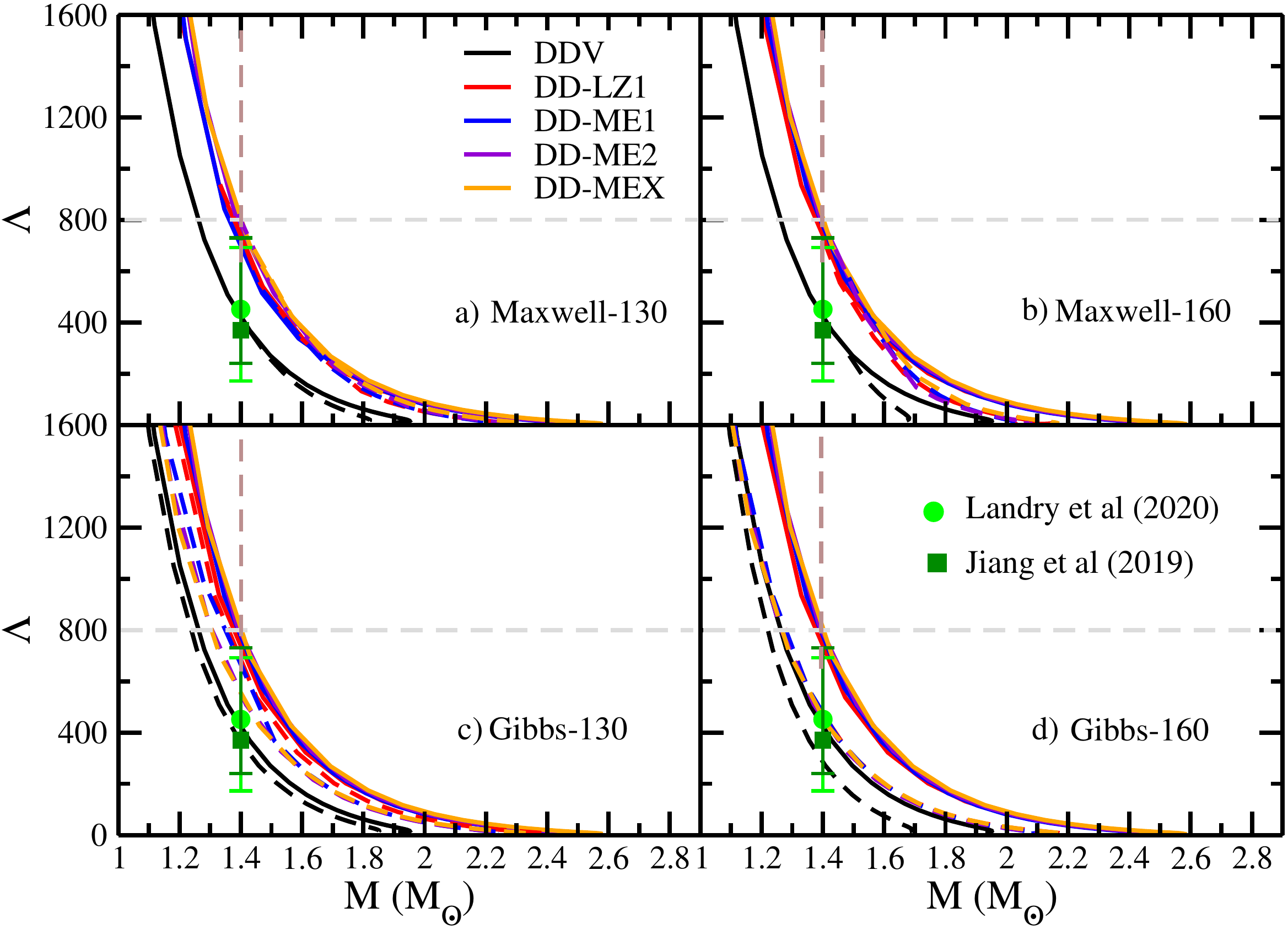}
	\caption{(color online) The dimensionless tidal deformability ($\Lambda$) as a function of NS mass corresponding to DDV, DD-LZ1, DD-ME1, DD-ME2, and DD-MEX EoSs and their HS configurations. The solid (dashed) lines represent MR plot for pure hadronic matter (hybrid NSs). The upper panels display the HS configuration with MC and the lower panels represent the same configuration with GC. The brown dashed line represents the NS canonical mass. The grey dashed line shows the upper limit of $\Lambda_{1.4}$ value from GW170817 \cite{PhysRevLett.119.161101}.  The non-parametric constraints on the tidal deformability of canonical NS mass are  shown\cite{PhysRevD.101.123007}. The constraint from joint PSR J0030+0451, GW170817, and the nuclear data analysis\cite{Jiang_2020}.}
	\label{fig5}
\end{figure}

To understand the configuration of the star produced by a given EoS model, the mas-radius profile is analyzed. The HS models are divided according to the phase transition construction and will be assessed by the pure hadronic configurations.

Fig.(\ref{fig4}) shows the mass-radius curves for pure hadronic DD-RMF EoSs (solid lines) and hybrid NS EoS (dashed lines) at different bag constants. The upper two panels represent the HS configuration with Maxwell construction at two different bag values $B_{eff}^{1/4}$=130 \& 160 MeV, while the lower panels represent the HS configuration with Gibbs construction at the same bag values 130 \& 160 MeV. The pure hadronic EoSs produce a NS with a maximum mass of $\approx$ 2.5$M_{\odot}$ for DD-LZ1, DD-ME1, DD-ME2, and DD-MEX parameter sets, while DDV set produces an NS with a maximum mass of 1.95 $M_{\odot}$. The vBag model parameters $K_{\nu}$ and $B_{eff}$ control the type of curves that result after the phase transition. While the $K_{\nu}$ parameter controls the stiffness of the curves,  $B_{eff}$ triggers the location of the phase transition along the curve.  In the upper two panels of fig. (\ref{fig4}), the MR curves produce a sharp discontinuous transition from hadron matter to quark matter due to sharp phase transition in Maxwell construction. The NS maximum mass is reduced from 2.555$M_{\odot}$ to 2.275$M_{\odot}$ for DD-LZ1 hybrid EoS at $B_{eff}^{1/4}$=130 MeV and further reduced to 2.146$M_{\odot}$ at $B_{eff}^{1/4}$=160 MeV. Other hybrid EoSS follow a similar pattern. The recent constraints on the maximum mass from PSR J1614-2230 (1.928$\pm$0.017$M_{\odot}$) \cite{Demorest2010}, PSR J0348+0432 (2.01$\pm$0.04$M_{\odot}$) \cite{Antoniadis1233232}, PSR J0740+6620              (2.14$^{+0.10}_{-0.09}$$M_{\odot}$) \cite{Cromartie2020}, and GW190814 (2.50-2.67 $M_{\odot}$) \cite{Abbott_2020a} are shown. The arrow represents the radius at the NS canonical mass with a maximum value $R_{1.4}\le13.76$km \cite{PhysRevLett.120.172702}. All the HS configurations satisfy the recently observed mass and radius constraints except that from DDV HS configuration whose maximum mass lies below the 1.9$M_{\odot}$ from PSR J1614-2230 at both 130 \& 160 MeV effective bag constant. This implies that a DDV EoS with phase transition to quark matter is too soft to satisfy the recent astrophysical constraints on the maximum mass and radius.\par

The HS models with Gibbs construction produce mass-radius curves with a smooth transition from hadron to quark matter because of the smoothly mixed phase in GC. The maximum mass of HS DD-LZ1 decreases from 2.555$M_{\odot}$ to 2.192$M_{\odot}$  at $B_{eff}^{1/4}$=130 MeV and to 2.043$M_{\odot}$ at $B_{eff}^{1/4}$=160 MeV. We see that the maximum mass in MC is higher than that for the GC case. It is because of the delayed phase transition in MC than the GC that allows the star to stay longer in the hadronic phase. The radius at the maximum mass changes from 12.297 to 12.475 km and 12.355 km at bag constant 130 \& 160 MeV, respectively. For MC, the radius changes to 12.428 and 12.574 km respectively. This shows that the Maxwell construction produces an NS with a large maximum mass and radius as compared to the Gibbs construction. However, the radius at the canonical mass, $R_{1.4}$, is same for MC as the pure hadronic star but smaller for GC as seen in the fig. (\ref{fig4}). Thus the maximum mass of the given NS configurations is lowered to satisfy the  2$M_{\odot}$ constraint.

\begin{center}
	\begin{table}[ht]
		\centering
		\caption{NS matter properties Maximum mass ($M_{max}$), corresponding radius ($R_{max}$),  canonical mass radius ($R_{1.4}$), and dimensionless tidal deformability ($\Lambda_{1.4}$) for pure hadron matter and HS configurations at effective bag constants $B_{eff}^{1/4}$=130 \& 160 MeV. The HS properties with both Gibbs as well as Maxwell construction are shown. }
		\vspace{0.2cm}
		\begin{tabular}{|c|c|cccccccc|}
			\hline
			\multirow{3}{*}{\makecell{Star \\ properties}} &\multirow{3}{*}{\makecell{Pure \\ Hadronic}}& \multicolumn{4}{c|}{Gibbs Construction} & %
			\multicolumn{4}{c|}{Maxwell Construction}\\
			\cline{3-10}
			& &\multicolumn{2}{c}{130 MeV} & \multicolumn{2}{c}{160 MeV} & \multicolumn{2}{c}{130 MeV} & \multicolumn{2}{c|}{160 MeV} \\
			\cline{3-10}
			& &\multicolumn{8}{c|}{DDV EoS}  \\
			\hline
			$M_{max}(M_{\odot})$&1.951&1.793&&1.665&&1.821&&1.680& \\
			$R_{max}$(km)&10.851&10.298&&10.805&&10.943&&11.211& \\
			$R_{1.4}$(km)&12.132&11.459&&11.427&&12.132&&12.132& \\
			$\Lambda_{1.4}$&392.052&356.261&&297.815&&392.052&&392.052&\\
			\cline{3-10}
			& &\multicolumn{8}{c|}{DD-LZ1 EoS}  \\
			\hline
			$M_{max}(M_{\odot})$&2.555&2.192&&2.043&&2.275&&2.146& \\
			$R_{max}$(km)&12.297&12.475&&12.355&&12.428&&12.574& \\
			$R_{1.4}$(km)&13.069&12.752&&12.706&&13.069&&13.069& \\
			$\Lambda_{1.4}$&728.351&698.233&&536.173&&728.351&&728.351&\\
			\cline{3-10}
			& &\multicolumn{8}{c|}{DD-ME1 EoS}  \\
			\hline
			$M_{max}(M_{\odot})$&2.449&2.106&&1.974&&2.219&&2.027& \\
			$R_{max}$(km)&11.981&11.211&&10.128&&12.162&&12.349& \\
			$R_{1.4}$(km)&12.898&11.507&&11.543&&12.898&&12.898& \\
			$\Lambda_{1.4}$&689.342&658.047&&495.146&&689.342&&689.342&\\
			\cline{3-10}
			& &\multicolumn{8}{c|}{DD-ME2 EoS}  \\
			\hline
			$M_{max}(M_{\odot})$&2.483&2.174&&2.008&&2.246&&2.074& \\
			$R_{max}$(km)&12.017&12.187&&12.013&&12.204&&12.391& \\
			$R_{1.4}$(km)&12.973&12.224&&12.247&&12.973&&12.973& \\
			$\Lambda_{1.4}$&733.149&572.844&&475.367&&733.149&&733.149&\\
			\cline{3-10}
			& &\multicolumn{8}{c|}{DD-MEX EoS}  \\
			\hline
			$M_{max}(M_{\odot})$&2.575&2.246&&2.095&&2.325&&2.164& \\
			$R_{max}$(km)&12.465&12.547&&12.506&&12.659&&12.754& \\
			$R_{1.4}$(km)&13.168&12.497&&12.451&&13.168&&13.168& \\
			$\Lambda_{1.4}$&791.483&594.376&&462.753&&791.483&&791.483&\\
			\hline
		\end{tabular}
		\label{tab4}
	\end{table}
\end{center}

Fig. (\ref{fig5}) displays the dimensionless tidal deformability as a function of NS mass for pure hadronic EoSs and HS configurations. As seen, the tidal deformability decreases with NS mass and becomes very small at the maximum mass. The softer EoS like pure hadronic DDV has tidal deformability at 1.4$M_{\odot}$, $\Lambda_{1.4}$=392.052, while for other stiffer EoSs, the value lies in the range $\Lambda_{1.4}$=690-790, which is well constrained by the upper limit on tidal deformability from GW170817 data \cite{PhysRevLett.119.161101}. The non-parametric constraints on the tidal deformability given by $\Lambda=451_{-279}^{+241}$ is shown\cite{PhysRevD.101.123007}. The constraints from the joint PSR J0030+0451, GW170817, and the nuclear data analysis at the canonical mass,$\Lambda_{1.4}=370_{-130}^{+360}$\cite{Jiang_2020} is also shown. For the phase transition with Maxwell construction, we see that the tidal deformability at the canonical mass remains the same as pure hadronic one, while for the Gibbs phase transition, the tidal deformability decreases from 728.351 to 698.233 and 536.173 for effective bag constants 130 \& 160 MeV, respectively. It is clear that the tidal deformability for the obtained HS configurations using the  Gibbs phase transition is favored over Maxwell construction by the tidal constraint from GW170817. We see that for the parameters used and phase transition construction, the maximum mass of each hybrid star curve is different from each other, which implies that no two curves produce a twin-star which share same maximum mass and hence show two branches of tidal deformability.

Table (\ref{tab4}) shows the NS properties like maximum mass, radius, radius at the canonical mass, and the dimensionless tidal deformability for pure hadronic phase and HSs. All the NS matter properties have been calculated for HS configurations using both Gibbs and Maxwell construction to see as to how the global and local charge neutrality conditions shape up the mass-radius of a NS. It is clear from the table that the radius at the canonical mass and the tidal deformability of an NS remains the same for Maxwell transition as that for pure hadronic matter, while both decrease for Gibbs transition. The obtained properties of HSs with Gibbs transition satisfy all the constraints from the recent observations of mass, radius, and tidal deformability. 

\begin{figure}[hbt!]
	\centering
	\includegraphics[width=10cm,height=8cm]{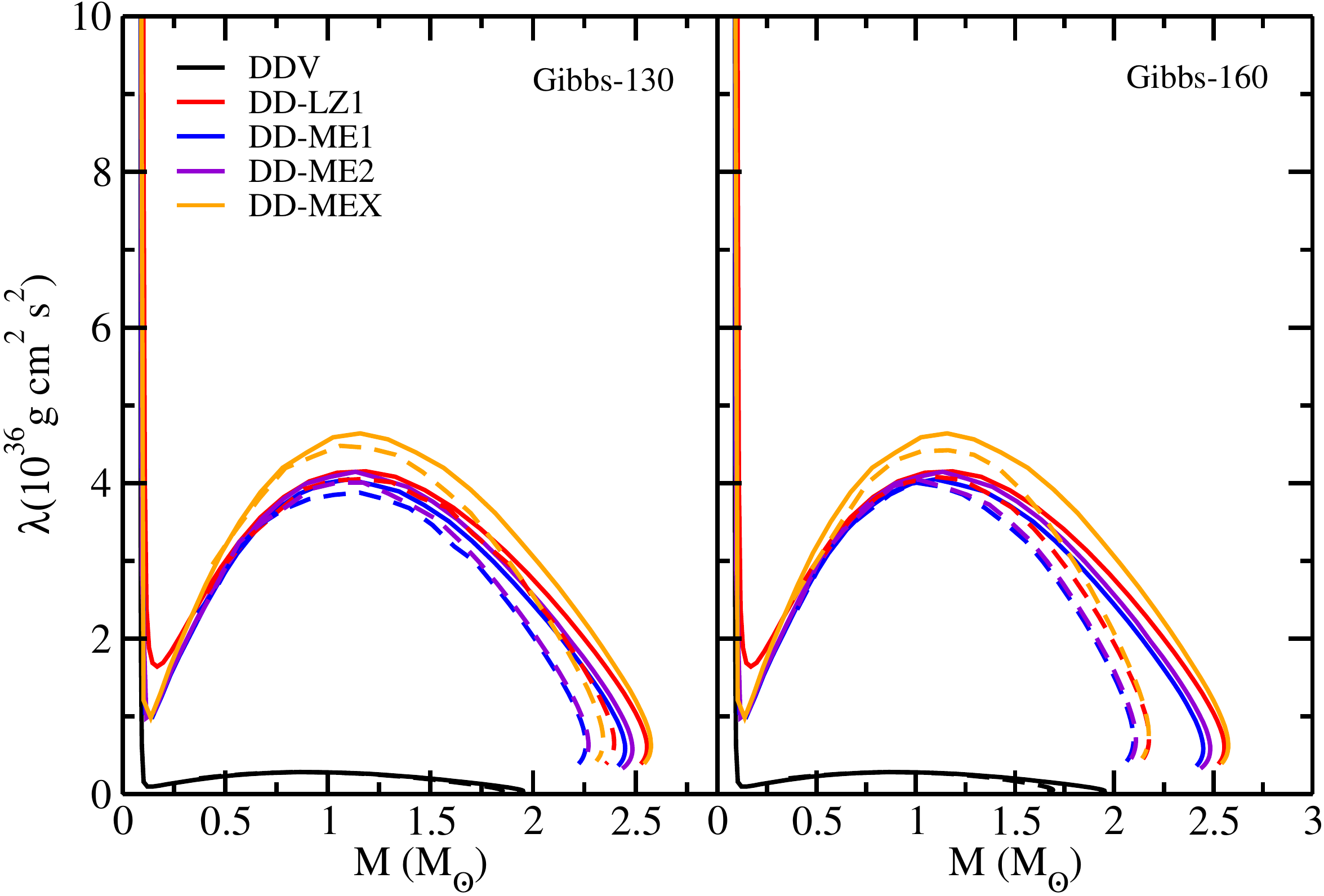}
	\caption{(color online) Tidal deformability $\lambda$ as a function of NS mass for pure hadronic and HS configurations. The solid (dashed) lines represent the pure hadronic (HS) EoSs. The left and right panel represents EoSs from Gibbs construction at  $B_{eff}^{1/4}$=130 MeV \& 160 MeV, respectively. }
	\label{fig6} 
\end{figure}
\begin{figure}[hbt!]
	\centering
	\includegraphics[width=10cm,height=8cm]{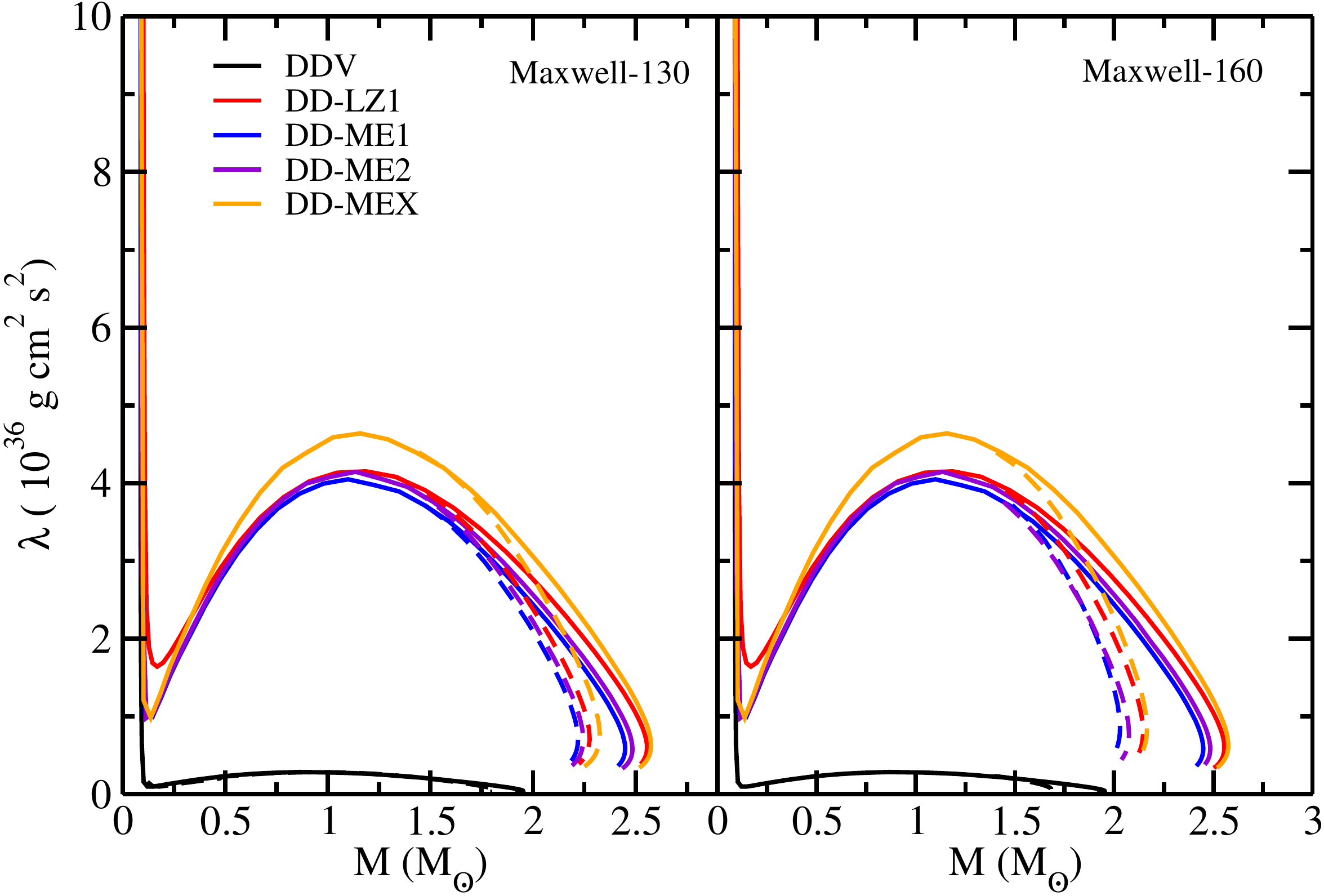}
	\caption{(color online) Same as fig.(\ref{fig6}), but for Maxwell phase transition at  $B_{eff}^{1/4}$=130 MeV (left panel) \& 160 MeV (right panel), respectively. }
	\label{fig7} 
\end{figure}

Fig. (\ref{fig6}) displays the dimensional tidal deformability as a function of NS mass for pure hadronic and HS configurations using  the Gibbs transition method. Apart from DDV parameter set, all other sets produce large values of tidal deformability and hence show large deformation. The shift in the tidal deformability to the lower values is seen in the figure. This shift is large for effective bag constant 160 MeV as compared to 130 MeV.

The tidal deformability vs NS mass plot is also shown in fig.(\ref{fig7}) for HSs with Maxwell phase transition to quark matter.

There is no shift in the tidal deformability for HS configurations at 1.4$M_{\odot}$. However the deviation at the NS maximum mass is visible. From eq.(\ref{l1}), the tidal deformability is very sensitive to the radius of the star and varies as $\lambda \propto R^5$. Thus the variation in the tidal deformability is directly affected by the nuclear EoS. DD-MEX set predicts a NS with large radius of 12.465 km due to its stiff nature at high density and thus gives large tidal deformability as compared to the other sets. DDV set produces very soft EoS and hence predicts a low value of tidal deformability.

\section{Summary and Conclusion}
\label{summary}
The recent detection of a black hole collision of mass 22.2-24.3 $M_{\odot}$ with a compact object of mass 2.50-2.67 $M_{\odot}$ by LVC was reported as GW190814. The secondary object of GW190814 attracted a lot of attention as it is either the heaviest neutron star or the lightest black hole ever discovered because of no tidal signatures or electromagnetic counterparts. We employed several latest density-dependent relativistic mean-field (DD-RMF) parameter sets like DDV, DD-LZ1, DD-ME1, DD-ME2, and DD-MEX to study the star matter properties. In order to study the phase transition from hadron matter to quark matter, the Vector-Enhanced Bag model (vBag) is employed for quark matter which accounts for the Dynamic Chiral Symmetry Breaking (D$\chi$SB) and repulsive vector interactions, explicitly. The hybrid star EoSs are generated by allowing a phase transition between hadron matter and quark matter. Both the Maxwell and Gibss methods are used to construct the mixed-phase between hadrons and quarks. The free parameter in the vBag model, $K_{\nu}$, which controls the stiffness of the EoS curve is fixed at $K_{\nu}=6$GeV$^{-2}$ for three flavour quark matter. The effective bag constant with values $B_{eff}^{1/4}$=130 \& 160 MeV are used. \\

By solving the TOV equation for the obtained pure and hybrid EoSs under $\beta$-equillibirum and charge-neutral conditions, the NS properties like mass, radius, and tidal deformability are calculated for all the configurations. The softer EoS DDV supports a NS with a maximum mass of 1.951$M_{\odot}$ at 10.851 km and tidal deformabilty at 1.4$M_{\odot}$, $\Lambda_{1.4}$=392.052. The presence of quarks lowers the maximum mass and the tidal deformability from 1.951 to 1.6 $M_{\odot}$ and 392 to around 297 using both Gibbs and Maxwell construction. For the stiffer EoSs like DD-LZ1, DD-ME1, DD-ME2, and DD-MEX, the NS maximum mass generated lies in the range 2.44-2.57$M_{\odot}$ which satisfies the mass constraint from GW190814 data. However, the phase transition between hadron and quark matter reduces the maximum to around 2$M_{\odot}$
which satisfies the constraints from the GW170817 data. The tidal deformability is also lowered from 790 to around 500. The Gibbs construction predicts a slightly lower maximum mass and the corresponding radius of NSs as compared to the Maxwell construction because of the delayed phase transition in the Maxwell method. While the tidal deformability of the HSs remains the same as that of pure hadronic stars in maxwell construction, it decreases with an increase in the bag constant for Gibbs construction. \\

Thus we see that the hadronic EoSs obtained using the recent DD-RMF parameterizations satisfy the mass constraint from GW190814 data, thus allowing us to consider the possibility of the secondary component of GW190814 as a massive NS. The phase transition from hadron matter to quark matter lowers the NS properties like mass, radius, and tidal deformability to satisfy the constraints from GW170817 data providing additional constraints on the NS maximum mass and hence on the dense matter EoS. The HSs produced via phase transition satisfy the 2$M_{\odot}$ limit. Therefore, the secondary component of GW190814 behaves as a massive NS and a phase transition to the quark matter allows it to become a HS. A more precise measurement of the NS maximum mass and the tidal deformability by the gravitational wave detectors will help improve and provide the proper EoS in the future.

\section*{Acknowledgement}
IAR is thankful to V. Dexheimer for detailed discussions and suggestions on the hadron quark phase transition.
	A.A.U. acknowledges the Inter-University Centre for Astronomy and Astrophysics, Pune, India for support via an associateship and for hospitality.

\section*{References}
\providecommand{\newblock}{}

\end{document}